\documentclass[letterpaper,twocolumn,showpacs,amsmath,amssymb,prl,superscriptaddress,groupedaddress]{revtex4}
\usepackage{graphicx}
\usepackage{dcolumn}
\usepackage{bm}
\usepackage{macros}

\preprint{APS/123-QED}

\begin{document}
    
\title{A New Limit on Time-Reversal Violation in Beta Decay}

\author{H.P.~Mumm}\affiliation{National Institute of Standards and Technology, Gaithersburg, MD 20899}\affiliation{CENPA and Physics Department, University of Washington, Seattle, WA 98195}
\author{T.E.~Chupp}\affiliation{Physics Department, University of Michigan, Ann Arbor, MI, 48104}
\author{R.L.~Cooper}\affiliation{Physics Department, University of Michigan, Ann Arbor, MI, 48104}
\author{K.P.~Coulter}\affiliation{Physics Department, University of Michigan, Ann Arbor, MI, 48104}
\author{S.J.~Freedman}\affiliation{Physics Department, University of California at Berkeley and Lawrence Berkeley National Laboratory, Berkeley, CA 94720}
\author{B.K.~Fujikawa}\affiliation{Physics Department, University of California at Berkeley and Lawrence Berkeley National Laboratory, Berkeley, CA 94720}
\author{A.~Garc\'{\i}a}\affiliation{CENPA and Physics Department, University of Washington, Seattle, WA 98195}\affiliation{Department of Physics, University of Notre Dame, Notre Dame, IN 46556}
\author{G.L.~Jones}\affiliation{Physics Department, Hamilton College, Clinton, NY 13323}
\author{J.S.~Nico}\affiliation{National Institute of Standards and Technology, Gaithersburg, MD 20899}
\author{A.K.~Thompson}\affiliation{National Institute of Standards and Technology, Gaithersburg, MD 20899}
\author{C.A.~Trull}\affiliation{Physics Department, Tulane University, New Orleans, LA 70118}
\author{J.F.~Wilkerson}\affiliation{Department of Physics and Astronomy, University of North Carolina, Chapel Hill, NC 27599}\affiliation{CENPA and Physics Department, University of Washington, Seattle, WA 98195}
\author{F.E.~Wietfeldt}\affiliation{Physics Department, Tulane University, New Orleans, LA 70118}

\date{\today}
\begin{abstract}
We report the results of an improved determination of the triple correlation  $D{\bf P}\cdot(\bf{p}_{e}\times\bf{p}_{v})$ that can be used to limit possible time-reversal invariance in the beta decay of polarized neutrons and constrain extensions to the Standard Model.  Our result is $D=(-0.96\pm 1.89 (stat)\pm 1.01 (sys))\times 10^{-4}$.  The corresponding phase between $g_A$ and $g_V$ is $\phi_{AV} = 180.013^\circ\pm0.028^\circ$ (68\% confidence level).  This result represents the most sensitive measurement of $D$ in nuclear beta decay.  
\end{abstract}

\pacs{24.80.1y, 11.30.Er, 12.15.Ji, 13.30.Ce}

\maketitle

The existence of charge-parity (CP) symmetry violation in nature is particularly important in that it is necessary to explain the preponderance of matter over antimatter in the universe~\cite{SAK91}. Thus far, CP violation has been observed only in the K and B meson systems~\cite{CHR64,AUB01,ABE01} and can be entirely accounted for by a phase in the Cabbibo-Kobayashi-Maskawa matrix in the electroweak Lagrangian. This phase is insufficient to account for the known baryon asymmetry in the context of Big Bang cosmology~\cite{RIO99}, so there is good reason to search for CP violation in other systems. As CP and time-reversal (T) violation can be related to each other through the CPT theorem, experimental limits on electric dipole moments and T-odd observables in nuclear beta decay place strict constraints on some, but not all, possible sources of new CP violation. 
\begin{figure*}[t]
       \includegraphics[width=6.25in]{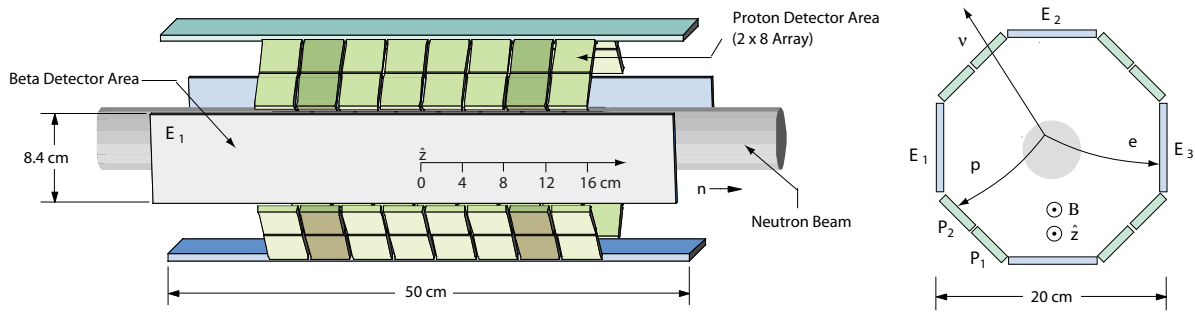}
        \caption{A schematic of the emiT detector illustrating the alternating electron and proton detector segments.  The darker shaded proton detectors indicate the the paired-ring at $z = \pm10$ cm. The cross section view illustrates, in an exaggerated manner, the effect of the magnetic field on the particle trajectories and average opening angle. A P$_2$E$_3$ coincidence event is shown.}
        \label{fig:DetGeom}
\end{figure*}

The decay probability distribution for neutron beta decay, $dW$, can be written in terms of the beam polarization $\bf{P}$ and the momenta (energies) of the electron ${\bf p}_e$ ($E_e$) and antineutrino ${\bf p}_\nu$ ($E_\nu$) as~\cite{JAC57} 
\begin{eqnarray}
    \label{eq:Jackson}
        \lefteqn{dW \propto 1+ a\frac{{\bf p}_{e}\cdot{\bf p}_{\nu}} {E_{e}E_{\nu}}+ b\frac{{m}_{e}} {E_{e}} +}&&\nonumber\\
        &&\hspace{.5 in}{\bf P} \cdot \left( A\frac{{\bf p}_{e}} {E_{e}}+ B\frac{{\bf p}_{\nu}} {E_{\nu}} + D\frac{{\bf p}_{e} \times {\bf p}_{\nu}}{E_{e}E_{\nu}} \right).
\end{eqnarray}

\noindent A contribution of the parity-even triple correlation $D{\bf P}\cdot({\bf p}_{e} \times {\bf p}_{\nu})$ above the level of calculable final-state interactions (FSI) implies T-violation.  
The PDG average of recent measurements is $D=(-4\pm6)\times 10^{-4}$~\cite{PDG:2010, LIS00, SOL04}, while the FSI for the neutron are $\sim10^{-5}$~\cite{CAL67,Ando2009}.  
Complementary limits can be set on other T-violating correlations, and recently a limit on $R$ has been published~\cite{Kozela:2009ni}. 
 Various theoretical models that extend the SM, such as left-right symmetric theories, leptoquarks, and certain exotic fermions could give rise to observable effects that are as large as the present experimental limits~\cite{HER98}.  Calculations performed within the Minimal Supersymmetric Model, however,  predict $D\lesssim 10^{-7}$~\cite{Drees2003}.

In the neutron rest frame, the triple correlation can be expressed as $D{\bf P} \cdot ({\bf p}_p \times {\bf p}_e)$, where ${\bf p}_p$ is the proton momentum.   
Thus one can extract $D$ from the spin dependence of proton-electron coincidences in
 the decay of cold polarized neutrons.
Our measurement was carried out at the National Institute of Standards and Technology Center for Neutron Research~\cite{MUM04}. The detector, shown schematically in Fig.~\ref{fig:DetGeom}, consisted of an octagonal array of four electron-detection planes and four proton-detection planes concentric with a longitudinally polarized beam.  The beam, with a neutron capture fluence rate at the detector of $1.7\times10^8$ cm$^{-2}$ s$^{-1}$, was defined using a series of \LiF\ apertures and polarized to $>91\%$ (95\% C.L.) by a double-sided bender-type supermirror~\cite{MUM04}. A 560 $\mu$T guide field maintained the polarization direction throughout the fiducial volume and a current-sheet spin-flipper was used to reverse the neutron spin direction every 10~s.  The symmetric octagonal geometry was chosen to maximize sensitivity to $D$ while approximately canceling systematic effects stemming from detector efficiency variations or coupling to the spin correlations $A$ and $B$~\cite{WAS94, LIS00}. 
Each of the four proton segments consisted of a $2\times8$ array of silicon surface-barrier detectors (SBDs) with an active layer 300\,mm$^2$$\times$ 300\,$\mu$m.  Each SBD was contained within an acceleration and focusing cell consisting of a 94\% transmitting grounded wire-mesh box through which the recoil protons entered.  Each SBD, situated within a field-shaping cylindrical tube, was held at a fixed voltage in the range $-25$\,kV to $-32$\,kV. The sensitive regions of the beta detectors were plastic scintillator measuring 50\,cm by 8.4\,cm by 0.64\,cm thick, with photomultiplier tube (PMT) readout at both ends. This thickness is sufficient to stop electrons at the decay endpoint energy of 782\,keV. 
The proton and beta detectors were periodically calibrated {\it in situ} with gamma and beta sources respectively. Details of the apparatus are presented elsewhere~\cite{MUM03,MUM04, LIS00}.

Data were acquired in a series of runs from October 2002 through December 2003.    Typical count rates were 3\,s$^{-1}$ and 100\,s$^{-1}$ for single proton and beta detectors, respectively, while the coincidence rate for the entire array was typically 25\,s$^{-1}$. 
 Of the raw events, 12\% were eliminated by filtering on various operational parameters (e.g. coil currents) and by requiring equal counting time in each spin-flip state.      
A beta-energy software threshold of 90~keV eliminated detection efficiency drifts due to changes in PMT gain coupled with the hardware threshold.
This was the largest single cut, eliminating 14\% of the raw events.  A requirement that a single beta be detected in coincidence with each proton eliminated 7\% of events. All cuts were varied to test for systematic effects.
\begin{figure}[t]
        \includegraphics[width=3.5 in]{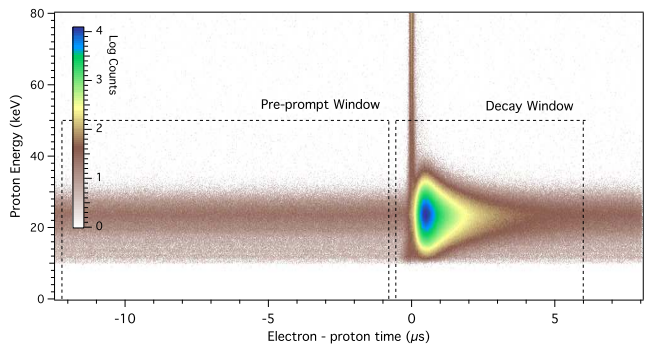}
        \caption{Intensity log plot of SBD-scintillator coincidence data showing proton energy vs delay time. Events near $\Delta t=0$ are prompt coincidences due primarily to beam-related backgrounds.
        }
        \label{fg:TOF}
\end{figure}

The remaining coincidence events were divided into two timing windows: a preprompt window from -12.3 $\mu$s to -0.75 $\mu$s that was used to determine the background from random coincidences, 
 and the decay window from -0.5 $\mu$s to 6.0 $\mu$s as shown in Fig.~\ref{fg:TOF}.  The recoil proton has an endpoint of 750~eV.  On average it is delayed by $\sim$ 0.5 $\mu$s.  
The average signal-to-background ratio was $\sim$ 30/1.  
 The energy-loss spectrum produced by minimum ionizing particles in 300 $\mu$m of silicon is peaked at approximately 100 keV and, being well separated from the proton energy spectrum, yielded an estimated contamination below 0.1\%.  
 The final data set consisted of approximately 300 million accepted coincidence events.




A detailed Monte Carlo simulation was used  to estimate a number of systematic effects.  The program \penelope{}~\cite{penelope}, which has been tested against data in a variety of circumstances of relevance to neutron decay~\cite{martin06},  was embedded within a custom tracking code.  All surfaces visible to decay particles were included.  The Monte Carlo was based on the measured beam distribution upstream and downstream of the fiducial volume~\cite{MUM04} and incorporated the magnetic field and electron energy threshold.  A separate Monte Carlo based on the package SIMION~\cite{SIMION}, incorporating the detailed geometry of the proton cells, was used to model the proton detection response function.




Achieving the desired sensitivity to $D$ in the presence of the much larger spin-asymmetries due to $A$ and $B$  depends critically on the measurement symmetry.  To the extent that this symmetry is broken, corrections must be applied to the measured result.  These corrections are listed in Table~\ref{tb:Systematics} and are discussed below.  To extract $D$, coincident events were first combined into approximately efficiency-independent asymmetries

\begin{equation}
\label{eqn:wdef}
w^{p_i e_j}=\frac{N^{p_i e_j}_+-N^{p_i e_j}_-}{N^{p_i e_j}_++N^{p_i e_j}_-},
\end{equation}

\noindent where $N^{p_ie_j}_+$ is the integrated number of coincident events in proton detector $i = 1...64$, beta detector $j=1...4$, with neutron spin + ($-$) aligned (anti-aligned) with the guide field.  For uniform polarization, ${\bf P}$, the asymmetries, $w^{p_ie_j}$, can be written in terms of decay correlations as

\begin{equation}
\label{wbyKdef}
w^{p_i e_j}\approx {\bf P}\cdot(A\tilde{\bf K}^{p_i e_j}_A +B\tilde{\bf K}^{p_i e_j}_B+D\tilde{\bf K}^{p_i e_j}_D),
\end{equation}

\noindent where the $\bf K$'s are obtained from Eqn.~\ref{eq:Jackson} by integrating the normalized kinematic terms over the phase space of the decay, the neutron beam volume, and the acceptance of the indicated detectors~\cite{LIS00}.  $\tilde{\bf K}_A \propto \langle{\bf p}_e/E_e\rangle$ and $\tilde{\bf K}_B\propto \langle{\bf p}_\nu/E_\nu\rangle$ are primarily transverse to the detector axis but have roughly equal longitudinal components for coincidence events involving the two beta detectors opposite from the indicated proton detector (E$_2$ and E$_3$ for P$_2$ as shown in Fig.~\ref{fig:DetGeom}).  The $\tilde{\bf K}_D$'s, however, are primarily along the detector axis and are opposite in sign for the two beta detectors.  Thus for each proton detector we can choose an appropriate combination of detector pairs that is sensitive to the $D$-correlation but that largely cancels the parity-violating $A$ and $B$ correlations.  One such combination is

\begin{equation}
\label{eqn:vdef}
v^{p_i}=\frac{1}{2}(w^{p_i e_R} - w^{p_i e_L}),
\end{equation}

\noindent where $e_R$ and $e_L$ label the electron-detector at approximately 135$^\circ$ giving a positive and negative cross-product ${\bf p}_p\times {\bf p}_e$ respectively (P$_2$E$_3$ vs P$_2$E$_2$ as shown in Fig.~\ref{fig:DetGeom}).
Proton cells at the detector ends accept decays with larger longitudinal components of $\tilde{\bf K}_A$ and are more sensitive to a range of effects that break the detector symmetry. 
We therefore define $\bar{v}$ as the average of the values of $v$ from the sixteen proton-cells  at the same $|z|$, i.e. $z = \pm$2, $\pm$6, $\pm$10, and $\pm$14 cm.  Each set of detectors corresponds to a paired-ring with the same symmetry as the full detector, e.g. the shaded detectors in Fig.~\ref{fig:DetGeom}.  We then define 
 
 \begin{equation}
\tilde D =  \frac{\bar{v}}{P\bar K_D},
\label{eq:tildeD}
  \end{equation}
  
\noindent where $\bar K_D = 0.378$ is the average of ${\hat z}\cdot(\tilde{\bf K}^{p_i e_R}_D - \tilde{\bf K}^{p_i e_L}_D)$ determined by Monte Carlo. The experiment provides four independent measurements  corresponding to each of the four paired-rings.  The systematic corrections to $\tilde D$ presented in Table I  yield our final value for $D$. 
 Eqn.~\ref{eq:tildeD} is based on the following:  1) accurate background corrections, 
2)  uniform proton and electron detection efficiencies, 3) cylindrical symmetry of the neutron beam and polarization, and 4) accurate determination of $\bar K_D$, $P$, and spin state. 

\begin{table}[t]
\small
\caption{Systematic corrections and combined standard uncertainties (68\% confidence level).
Values should be multiplied by $10^{-4}$.
\label{tb:Systematics}}
\begin{center}
\begin{tabular}{lll}
\hline
Source 								& Correction  		\hspace{0.02 in}	& Uncertainty 		\\
\hline\hline
Background asymmetry	\hspace{0.2 in}				& \hspace{0.1 in}$ 0$$^{\rm a}$					& $0.30$			\\
Background subtraction					&  \hspace{0.1 in}$0.03$	\hspace{0.18 in}	       		&$ 0.003$ 		\\
Electron backscattering			 		&  \hspace{0.1 in}$0.11$							&  $0.03$			\\
Proton backscattering 					&  \hspace{0.1 in}$0$$^{\rm a}$					&  $0.03$			\\
Beta threshold 							& \hspace{0.1 in}$0.04$							&  $0.10$			\\
Proton threshold						&$-0.29$ 							& $0.41$			\\
Beam expansion, magnetic field \hspace{0.12 in}& $-1.50$						& $0.40$			\\
Polarization non-uniformity				&  \hspace{0.1 in}$ 0$$^{\rm a}$					& $0.10$			\\
ATP - misalignment 						& $-0.07$							& $0.72$			\\
ATP - Twist 							& \hspace{0.1 in}$0$$^{\rm a}$					& $0.24$			\\
Spin-correlated flux$^{\rm b}$				& \hspace{0.1 in}$0$$^{\rm a}$					& $3\times 10^{-6}$	\\
Spin-correlated pol.						&  \hspace{0.1 in}$0$$^{\rm a}$						&$5\times 10^{-4}$	\\
Polarization$^{\rm c}$					&  $$								& $0.04$$^{\rm d}$  	\\
$\bar K_D$$^{\rm c}$					& $$								& $0.05$			\\
\hline
Total systematic corrections				& $-1.68$							& $1.01$			\\
\end{tabular}
\end{center}
\raggedright
\vspace{0.05 in}
$^{\rm a}$ Zero indicates no correction applied.  $^{\rm b}$ Includes spin-flip time, cycle asymmetry, and flux variation. $^{\rm c}$ Included in the definition of $\tilde D$. $^{\rm d}$ Assumed polarization uncertainty of 0.05.  \\
\end{table}




Backgrounds not properly accounted for contribute two systematic errors:  1) multiplicative errors due to dilution of the asymmetries, and 2) spin-dependent backgrounds 
 that can lead to a false $D$. 
 Errors in background subtraction, as well as possible spin-dependent asymmetries in this background, have a small effect. 
The multiplicative correction to the value of $w^{p_i e_j}$ due to backscattered electrons was determined using the Monte Carlo. 
The uncertainty given in Table~\ref{tb:Systematics} reflects the 20\%  uncertainty assigned to the backscattering fractions  due to  limitations of the detector and beam model and due to limited knowledge of backscattering at energies below a few hundred keV. Proton backscattering, though observable, produces a negligible effect on $\tilde{D}$. 

In principle, the values of $w^{p_i e_j}$ are independent of the absolute efficiencies of the proton and electron detectors; however, they do depend on any energy dependence of the efficiencies through the factors $\langle{\bf p}_e/E_e\rangle$. 
Spatial variation of the efficiencies breaks the symmetry assumed in combining proton-cell data into paired-rings. Beta energy thresholds were observed to vary less than 20 keV across the detector, implying the almost negligible correction given in Table~\ref{tb:Systematics}. Proton detector efficiency variations, however, were more significant.  Lower energy proton thresholds varied across the detector and over the course of the experiment.
These thresholds combined with  the spin-dependence of the accelerated proton energy spectra can result in 
significant deviations in the values of $w^{p_i e_j}$, though the effect on the the value of $v^{p_i}$ is largely mitigated because the low-energy portion of the proton energy spectrum is roughly the same for the $e_R$ and $e_L$ coincidence pairs. 
To estimate the proton-threshold-nonuniformity effect on $\tilde D$, spin-dependent proton energy spectra were generated by Monte Carlo for all proton-detector-electron-detector pairings and convoluted with model detector response functions based on fits to the average proton-SBD spectra.  The average fit parameters were varied over a range characteristic of the observed variations during the run.
Representative thresholds were then applied to determine the effect on $\tilde D$.  An alternative and consistent estimate was derived by correcting the values of $w^{p_i e_j}$ on a day-by-day basis using the spectrum centroid shift and empirically determined functional form of the spectrum at the threshold.

\begin{figure}[htpb]
\begin{centering}
\includegraphics[width=3.4 truein]{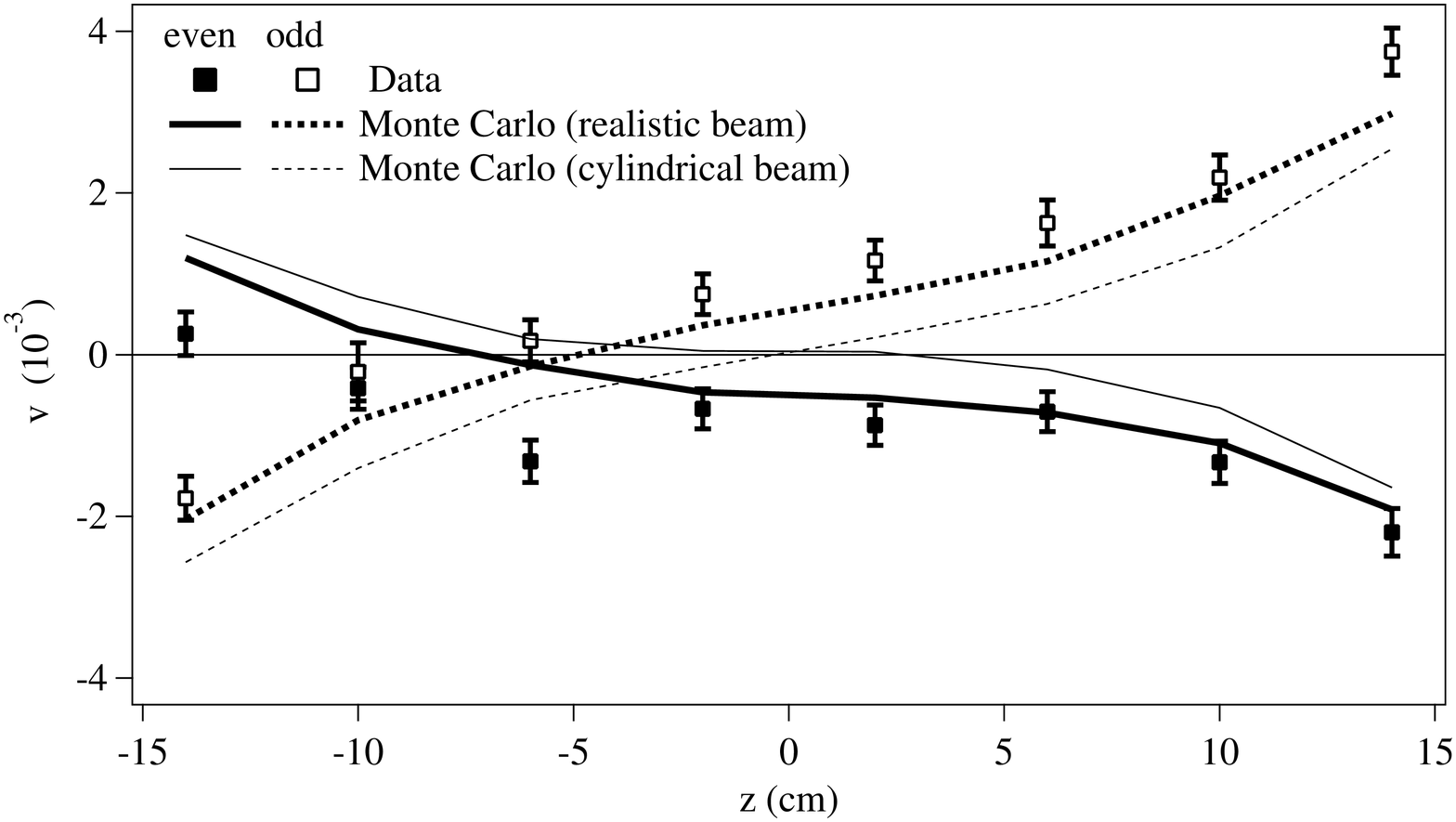}
\caption{Solid (open) squares show the values of $v$ averaged over the four planes for proton cells on the even (odd) side of the proton detection plane (the side with P$_2$ vs P$_1$ in Fig.~\ref{fig:DetGeom}).  Monte Carlo results are indicated by lines. The broken symmetry, due to the combination of the magnetic field and beam expansion, is evident in the shift in the crossing point from the detector center.
\label{fg:BeamShape}}
\end{centering}
\end{figure}

Beam expansion from a radius of 2.5~cm to 2.75~cm combined with the magnetic field breaks the symmetry of the detector because the average proton-electron opening angle for each proton-electron detector pair is modified.  Monte Carlo calculations using measured upstream and downstream density profile maps were used to calculate the correction given in Table~\ref{tb:Systematics}. Possible inaccuracies in the determination of the beam density were estimated and their implications explored with the Monte Carlo code.  
The effect on the value of $v^{p_i}$ as a function of position is illustrated in Fig.~\ref{fg:BeamShape}.  

For a symmetric beam, contributions to $v^{p_i}$ due to transverse polarization cancel for opposing proton planes; however small azimuthal beam asymmetries can affect this cancellation.  This asymmetric-beam transverse-polarization (ATP) effect is proportional to both $\sin\theta_P$, the angle of the average neutron-spin orientation with respect to the detector axis, and $\sin(\phi_P-\phi^{p_i})$, where  $\phi^{p_i}$ is the effective azimuthal position of the proton cell, and $\phi_P$ is the azimuthal direction of the neutron polarization. 
To study this effect transverse-polarization calibration runs with $\theta_P=90^\circ$ and several values of $\phi_P$ were taken over the course of the experiment.  In these runs the ATP effect was amplified by $\approx 200$.  
The values of $\sin \theta_P$ and $\phi_P$ for the experiment were determined using the calibration runs and Monte Carlo corrections for the beam density variations.
To estimate the effect, the extreme value of sin$\theta_P= 12.8\times10^{-3}$ and the range of $-31.5^\circ < \phi_P < 112.2^\circ$ were used. The uncertainty is due to uncertainties in the angles $\theta_P$ and $\phi_P$.  The effect of nonuniform beam polarization is also given in Table~\ref{tb:Systematics}. Time-dependent variations in flux, polarization, and the spin-flipper, as well as the uncertainty in the instrumental constant $\bar K_D$, can be shown to produce asymmetries proportional to $\tilde D$.  These effects are listed in Table~\ref{tb:Systematics}.



\begin{figure}[htdp]
\includegraphics[width=3.3in]{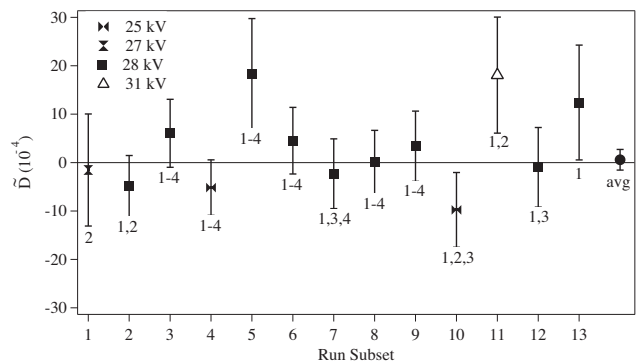}
\caption{Results for $\tilde D$ by run subset. Uncertainties are statistical. The fully functioning paired-rings used for each subset are indicated: 1-4 indicates all paired-rings were used.}
\label{fg:DSuperSliceSubsets}
\end{figure}

Correlations of $\tilde D$ with a variety of experimental parameters were studied  by varying the cuts and by breaking the data up into subsets taken under different conditions of proton acceleration voltage and number of live SBDs as shown in  Fig.~\ref{fg:DSuperSliceSubsets}.
  A linear correlation of $\tilde D$ with high-voltage, revealed by the cuts study, yields $\chi^2 = 5.6$ with 11 DOF compared to 10.4 for 12 DOF for no correlation.  The acceleration-voltage dependence of the focusing properties was extensively studied by Monte Carlo with no expected effect, and we intrepret the 2.1 sigma slope as a statistical fluctuation.

To improve the symmetry of the the detector, we combine counts for the entire run to determine the values of $v^{p_i}$ and to extract $\tilde D$ for each paired-ring. A blind analysis was performed by adding a constant hidden factor to Eqn.~\ref{eqn:wdef}.  The blind was only removed once all analyses of systematic effects were complete and had been combined into a final result. The weighted average of $\tilde D$ for the four paired-rings is  $0.72\pm 1.89$ with $\chi^2=0.73$ for 3 DOF.

The result including all corrections from Table~\ref{tb:Systematics} is
$$
D=(-0.96\pm 1.89 (stat)\pm 1.01 (sys))\times 10^{-4}.
$$

\noindent Our result represents the most sensitive measurement of the $D$ coefficient in nuclear beta decay.
Assuming purely vector and axial-vector currents, $\phi_{AV} = 180.013^\circ\pm0.028^\circ$ which is the best direct determination of a possible CP-violating phase  between the axial and vector currents in nuclear beta decay. Previously the most sensitive measurement was in $^{19}$Ne, with $D=(1\pm6)\times 10^{-4}$~\cite{CAL85}.


The authors acknowledge the support of the National Institute of Standards and Technology, U.S. Department of Commerce, in providing the neutron facilities used in this work. This research was made possible in part by grants from the U.S. Department of Energy Office of Nuclear Physics (DE-FG02-97ER41020, DE-AC02-05CH11231,
 and DE-FG02-97ER41041) and the National Science Foundation (PHY-0555432, PHY-0855694, PHY-0555474, and PHY-0855310). 

\bibliographystyle{prsty}
\bibliography{emiTRef}

\end{document}